%% file: main.tex
\documentclass[prd,amsfonts,onecolumn,superscriptaddress,aps,nofootinbib,11pt]{revtex4-1}

\usepackage{amsmath, amsthm, amssymb, amsfonts, mathrsfs}
\usepackage{mathtools}
\usepackage{braket, slashed, bm}
\usepackage{hyperref}
\usepackage{booktabs}
\usepackage{scalefnt}
\usepackage{setspace}
\usepackage{amsmath, amsthm, amssymb, amsfonts, mathrsfs}
\usepackage{scalefnt}
\usepackage{physics, braket, slashed, bm}
\usepackage{bbold}
\usepackage{listings}
\usepackage{siunitx}
\usepackage{booktabs}
\usepackage{soul}
\usepackage{color}
\usepackage[usenames,dvipsnames]{xcolor}
\usepackage{graphicx}
\usepackage{float}
\usepackage{placeins}
\usepackage{booktabs}
\usepackage{hyperref}
\usepackage[compat=1.1.0]{tikz-feynman}
\usepackage[caption=false]{subfig}
\usepackage{soul}
\usepackage{ulem}

\usepackage[compat=1.1.0]{tikz-feynman}

\hypersetup{
	unicode=false,          
	pdftoolbar=true,        
	pdfmenubar=true,        
	pdffitwindow=false,     
	pdfauthor={William},     
	colorlinks=true,       
	linkcolor=blue,          
	citecolor=magenta,        
	urlcolor=blue
}

\begin{document}

\title{Exploring Top-Quark Signatures of Heavy Flavor-Violating Scalars at the LHC with Parametrized Neural Networks}

\author{Alexandre Alves}
\email{aalves@unifesp.br}
\affiliation{Departamento de F\'isica, Universidade Federal de S\~ao Paulo, UNIFESP, 09972-270, Diadema-SP, Brazil}
\author{Eduardo da Silva Almeida}
\email{almeidae@ufba.br}
\affiliation{Departamento de Física do Estado Sólido, Universidade Federal da Bahia,\\
UFBA, 40170-115, Salvador-BA, Brazil }
\author{Alex G. Dias}
\email{alex.dias@ufabc.edu.br}
\affiliation{Centro de Ci\^encias Naturais e Humanas, Universidade Federal do ABC,\\
UFABC, 09210-580, Santo Andr\'e-SP, Brazil}
\author{Diego S. V. Gon\c{c}alves}
\email{diego.vieira@ufabc.edu.br}
\affiliation{Centro de Ci\^encias Naturais e Humanas, Universidade Federal do ABC,\\
UFABC, 09210-580, Santo Andr\'e-SP, Brazil}

\date{\today}
	
\begin{abstract}
    In this work, we study flavor-violating scalars (flavons) in a range of large masses that have not been explored previously. We model the interactions with an effective field theory formulation where the flavon is heavier than the top quark. In addition, we assume that the flavon only couples to fermions of the Standard Model in a flavor-changing way. As the flavon couples strongly to top quarks, same-sign and opposite-sign top quark pair signals can be explored in the search for those particles. Using parametrized neural networks, we show that it is possible to probe flavons with masses in the 200--1600 GeV range through their interactions with a top quark plus up and charm quarks for effective couplings of order $10^{-2}$ TeV$^{-1}$ at the 14 TeV High-Luminosity LHC.
    
\end{abstract}

\maketitle 
	
\section{Introduction}

Hypothetical scalar particles with flavor-violating interactions are part of the particle spectra of several models proposed to tackle problems that do not find a solution within the Standard Model (SM). Among the models predicting such particles are the ones based on the Froggatt-Nielsen mechanism \cite{Froggatt:1978nt}, where the hierarchical flavor structure of the SM fermions originates from suppressed effective Yukawa couplings given by powers of the ratio of the condensation scale of a scalar field breaking the global flavor $U(1)_F$ symmetry -- one that acts differently in each fermion multiplet of the gauge symmetry -- and a high-energy scale. The particle excitation of the pseudo-Nanbu-Goldstone boson field of this symmetry breaking is the $\it{flavon}$ and has flavor-violating interactions. Axions and axion-like particles \cite{DiLuzio:2020wdo,Jaeckel:2010ni} can also have flavor-violating interactions when their associated symmetry acts similarly to a flavor symmetry \cite{Ema:2016ops,Bjorkeroth:2018dzu,Bauer:2021mvw,Jho:2022snj,Dias:2021lmf}. As is typical of these particles, their flavor-violating interactions with matter are proportional to the mass of the fermions involved but suppressed by the new physics energy scale. Especially for axion-like particles and flavons\footnote{In the typical, axion models the suppression scale is too high \cite{DiLuzio:2017pfr}. However, it is possible to constrain axion flavor interactions in meson decays \cite{Bjorkeroth:2018dzu}.}, the suppression is such that it would still allow for signals of their production at the LHC~\cite{Bauer:2017ris}.  

In this work, we investigate signals of heavy flavons at the LHC 14 TeV under the hypothesis that their main interactions violate flavor (the maximally violating flavons) with perturbative couplings proportional to the quark masses. We explore the effective top quark-flavon couplings which are not still severely constrained, different from the flavor-violating couplings involving the down-type quarks. We assume that flavor-conserving couplings of the flavon to the fermions, as well as to photons, gluons, and electroweak gauge boson pairs, are relatively small compared to the considered flavor-violating ones.  

Our analysis focuses on flavon masses higher than 200 GeV, which decay predominantly into a top quark plus an up/charm quark. We identify a peculiar phenomenology of these new particles where final states contain both same-sign and opposite-sign top quark pairs, impacting the signal/background separation. Same-charge tops signals from flavor-violating scalars in two-Higgs doublets models have been searched by ATLAS~\cite{ATLAS:2023tlp} and CMS~\cite{CMS:2023xpx}.  

Machine learning (ML) techniques \cite{Mehta:2018dln,Carleo:2019ptp,Bourilkov:2019yoi,Radovic:2018dip} are powerful tools in new physics searches where signal-to-background ratios are expected to be too small to produce statistically significant signals.
We chose to employ a parametrized neural network (pNN) \cite{pnn1,pnn2} for the signal-background classification task to boost the discovery and exclusion limits of our effective field theory (EFT). The pNN option is adequate if we want the neural networks to learn to separate signals from backgrounds as a continuous function of the flavon mass parameter. 

Our main result is that flavons in the considered mass interval can be discovered, or excluded, in the 14 TeV High-Luminosity LHC (HL-LHC) for effective couplings up to order $10^{-2}$ TeV$^{-1}$, extending considerably the search prospects for these particles. The findings of the present analysis represent a complementary study of previous works~\cite{Alves:2023sdf,Cornella:2019uxs} for flavons lighter than the top quark.

In section~\ref{sec:eft}, we present the effective field theory formulation of the interactions; in section~\ref{sec:simulations}, the signal and background simulations are discussed; the analysis strategy and the neural network training are discussed in section~\ref{sec:analysis}; our results are presented in section~\ref{sec:results}, while the conclusions can be found in section~\ref{sec:conclusions}.

\section{Effective Lagrangian}
\label{sec:eft}

We assume the flavon field originates from the spontaneous symmetry breaking of a global $U(1)_F$ flavor symmetry as in the Froggatt-Nielsen scheme \cite{Froggatt:1978nt}. 
The main interactions of the flavon with fermions arise from effective operators like $y_{ij}(\frac{\phi}{M})^{n_f}\bar{F}_{iL}\Phi f^\prime_{jR}$, with $y_{ij}$ couplings that in principle could be of order one; $M$ a high-energy scale; $\phi(x)=(\rho(x)+\sqrt2\langle \phi\rangle)e^{ia(x)/\Lambda}/\sqrt{2}$ a scalar singlet field of the SM, hosting the flavon field $a$ and having vacuum expectation value $\langle \phi\rangle$ breaking the $U(1)_F$ symmetry; $F_{iL}$ the fermion doublet of left-handed fields and $f^\prime_{jR}$ the right-handed singlet fermion fields; $\Phi=H$ (or $i\sigma_2 H^*$ depending on the fermion $f^\prime_{jR}$) the Higgs doublet, and $n_f$ integer numbers fixed by the flavor symmetry. The scale $\Lambda$ may depend on the vacuum expectation values of other scalar fields charged under $U(1)_F$\footnote{This is, in fact, the flavon decay constant and can depend on the $U(1)_F$ charge $c_i$ of the scalar fields $\varphi_i$ getting vacuum expectation value as $\Lambda=\sqrt{2(\sum_i c_i^2 \langle \varphi_i\rangle^2)}$.}. With the vacuum expectation values of $H$ and $\phi$, we have that the flavon interactions with SM  fermions are described by the effective Lagrangian (e.g.~\cite{Ema:2016ops}) 
\begin{equation}
\label{eq:LFV-lagrangian}
    \mathcal{L} \supset -i \frac{a}{\Lambda} \sum_{f, i, j} \overline{f_i} \left[ (m_j - m_i) v_{ij} +  (m_j + m_i) a_{ij} \gamma_5 \right] f_j \,.
\end{equation}
where $m_i$ is the mass of the mass eigenstate fermion field $f_i$; $a_{ij}$ and $v_{ij}$ the axial and vectorial coefficients, respectively.

Depending on the coefficients $a_{ij}$ and $v_{ij}$, given the mass proportionality in Eq. (\ref{eq:LFV-lagrangian}), the most intense interactions of the flavon would involve the charged fermions of the third generation. Since $a$ is a pseudoscalar field, the flavor-violating (FV) couplings $v_{ij}$ imply parity symmetry violation.  However, the CP symmetry is preserved if $v_{ij}$ and $a_{ij}$ are real, which is the case we will study. 

It is also supposed that the flavon has a mass that could be generated by some explicitly breaking mechanism of the $U(1)_F$ symmetry through an anomalous feature, as in the case of the axion, or directly by an potential term like $V=({\lambda}\phi^n/{M^{n-4}} +h.c.)$. For our purposes the origin of this mechanism is not relevant so that we simply assume a mass $m_a$ for the flavon.

Our analysis will be based on the following hypotheses concerning the couplings: the flavon diagonal couplings with fermions, vector bosons, as well as the non-diagonal couplings with the down-type quarks are all negligible; the dominant couplings with quarks are the non-diagonal ones involving the top quark plus charm or up quark; the couplings with leptons play a subdominant role as the flavon couples to the fermion mass. This way, the set of parameters we consider is the following: 
\begin{equation}
\left\{
    \begin{array}{cc}
m_a, \Lambda & \hbox{flavon mass and new physics scale} \\
v_{tc}, a_{tc}, v_{tu}, a_{tu} & \hbox{non-diagonal top-quark couplings} 
    \end{array}
    \right.
    \label{eq:condition}
\end{equation}
These couplings avoid dangerous flavor-changing neutral currents at tree level and strong experimental constraints on flavor-diagonal interactions.

To restrict the parameters space further, we will also assume universal flavon couplings to quarks, that is, $v_{tc}=v_{tu}\equiv v_{tq}$, and $a_{tc}=a_{tu}\equiv a_{tq}$. Moreover, the cross sections will depend only on the ratio between $c_{tq}\equiv \sqrt{v_{tq}^2+a_{tq}^2}$ and $\Lambda$, thus our searches can be performed on the $c_{tq}/\Lambda$ {\it versus} the flavon mass, $m_a$.

Our analysis is restricted to flavons with masses above 200 GeV, which prefer to decay into top plus up or charm quarks. Even though the effective flavon couplings to down-type quarks are assumed to be negligibly small, flavons can decay to $a \rightarrow bq$ with $q = \bar{d},\bar{s}$ at 1-loop. However, this decay is suppressed compared to the top-quark decay channel since it arises from next-to-next leading order corrections \cite{Alves:2023sdf}. 

\section{SIGNAL AND BACKGROUND SIMULATIONS}
\label{sec:simulations}

We simulated signals and backgrounds in leading order at the 14 TeV LHC with \texttt{MadGraph5}~\cite{Alwall:2014hca, Frederix:2018nkq} using the \texttt{UFO}~\cite{Degrande:2011ua} files generated by \texttt{FeynRules}~\cite{Alloul:2013bka}, where our EFT was implemented.  The parton shower and detector simulations were performed with default settings of \texttt{PYTHIA 8.3} \cite{Bierlich:2022pfr} and \texttt{Delphes3}~\cite{deFavereau:2013fsa}, respectively. 

\begin{figure}[t]
    \centering
    \input{qg_at_T} 
    \input{qq_aa} \input{at}
\caption{Feynman diagrams for top pair signals. The left and middle diagrams depict the contributions from single and double production of flavons, respectively, while the rightmost diagram shows the virtual flavon contribution in the $t$-channel. Here, $q=u,c$.}
    \label{fig:prod}
\end{figure}
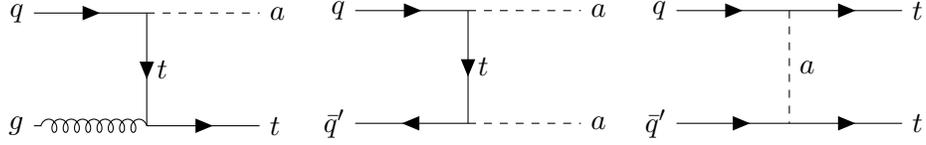
The signal we consider involves the production of 2 top quarks with the same or opposite charges and up to 2 non-$b$ jets after the decay of flavons into a top quark plus a $c$ or $u$ quark. We assume that flavons with masses higher than $m_t+m_c$ decay exclusively and equally into the dominating flavor-violating channels, $a\to t\bar{u}, t\bar{c}, \bar{t}u, \bar{t}c$.  The signal receives the following contributions:
\begin{itemize}
    \item[(a)] flavon in the $t$-channel, as shown at the right panel of Fig.~\ref{fig:prod}. In this case, two top quarks of any combination of charges are produced, $tt$, $\bar{t}\bar{t}$, $t\bar{t}$. After decay, the top quarks produce two $b$ jets and same-sign (SS) and opposite-sign (OS) leptons ($\ell=e,\mu$);
    \item[(b)] flavon single production, as shown in the left panel of Fig.~\ref{fig:prod}. Again, we have OS and SS top quark pairs but with a non-$b$ jet from the flavon decay;
    \item[(c)] double flavon production as shown in the middle panel of Fig.~\ref{fig:prod}. Now, we have OS and SS top quark pairs along with two non-$b$ jets from the flavon decay.
\end{itemize}

Summarizing, we propose to look for heavy flavons in the channels
\begin{eqnarray}
    && pp\to tt,\; \bar{t}\bar{t}+\hbox{jets}\to bb(\bar{b}\bar{b})+\ell^\pm\ell^\pm+\not\!\! E_T + \hbox{jets}\\
    && pp\to t\bar{t}+\hbox{jets}\to b\bar{b}+\ell^\pm\ell^\mp+\not\!\! E_T + \hbox{jets}\; .
\end{eqnarray}

\begin{figure}[t]
    \centering
    \includegraphics[width=0.45\linewidth]{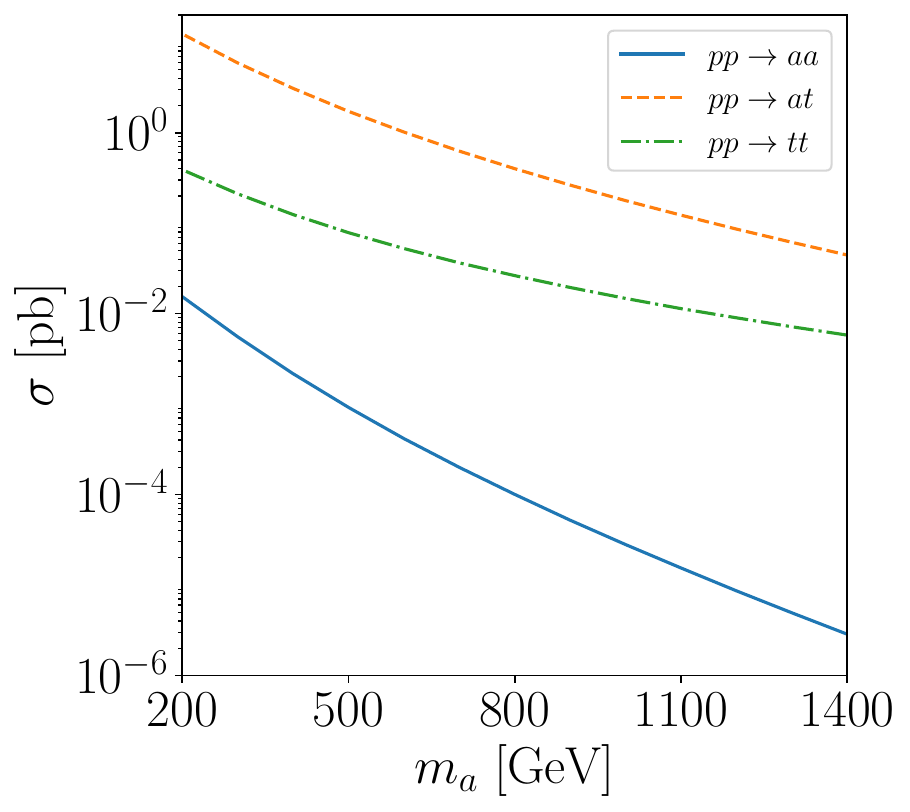}
    \caption{Production cross sections of all the three production modes considered in this work: pair production, $pp\to aa$; single production, $pp\to at$; $t$-channel contribution, $pp\to tt$. In this last case, all combinations of top quark charges were included.}
    \label{fig:xsec}
\end{figure}

 The production cross sections, in pb, for all three channels are shown in Fig.~\ref{fig:xsec}. For the range of masses we are interested in, the single production dominates, followed by the $t$-channel contribution and the double flavon production, which is suppressed by the mass of the flavon. 

 The main background sources for the OS leptons signals are $t\bar{t}$ as the dominating one, next-to-next-to-leading order Drell-Yan plus two $b$ jets ($\ell\ell bb$), and single top $tW$ as subdominant. Same-sign lepton signals have small reducible backgrounds from misidentifying leptons and jets from the three background sources described. The only irreducible background is the vector boson fusion (VBF) channel $qq'\to W^\pm W^\pm + bb$, which is CKM suppressed. Moreover, in this case, the bottom jets are produced in the forward regions of the detector, where $b$-tagging is less efficient, while the signals tend to be produced in the central regions of the detector. For these reasons, we neglect VBF events but collect all events with two SS leptons from the background sources described previously.

\section{ANALYSIS Strategy AND PARAMETRIZED NEURAL NETWORKs}
\label{sec:analysis}

\subsection{Basic Requirements}

Before training the neural networks, we impose kinematic cuts on both SS and OS signals and backgrounds. We exclude particles that do not meet the following selection
\begin{eqnarray}
\label{eq:basic-cuts}
    && \text{at least 1 $b$-jet} \quad \text{and} \quad \text{exactly 2 leptons}, \nonumber\\
    && p_{T_{\ell}} > 10\; \mbox{GeV},\; p_{T_{j,b}} > 20\; \mbox{GeV},\; \left|\eta_{\ell}\right| < 2.5,\; \left|\eta_{j,b}\right| < 2.5,\nonumber \\ 
    && M_{\ell\ell} > 100\;\mbox{GeV},\; \slashed{E}_T > 20\;\si{GeV}  \; .
\end{eqnarray}
Requiring at least one $b$-jet efficiently raises the overall $b$-tag to $\sim 70$\%. On the other hand, requiring two leptons drops the efficiency of signals, especially for heavy masses, as it gets more difficult to find isolated leptons. The cut on the invariant mass of two leptons, $M_{\ell\ell}$, helps to tame the large Drell-Yan background.

In Table~\ref{tab:cut-efficiencies}, we display the cross sections after imposing the cuts of Eq.~\eqref{eq:basic-cuts}. The background rates are still much larger than the signals. We could have selected events with same-sign leptons to clean up almost all those backgrounds at the cost of half of the signal events, but because it is easy to separate SS and OS events, we can use the OS events if we are able to classify signals versus backgrounds in this case.

\begin{table}[t]
    \centering
    \begin{tabular}{cccccc}
    \hline
          $t \bar{t}$ (pb)& $tW$ (pb)& $\ell\ell bb$ (pb) & $200$ GeV (fb) & $1000$ GeV (fb)& $1400$ GeV (fb)\\
          \hline\hline
           7.3& 0.76& 0.07 & 45.1 & 1.1 & 0.2 \\
          \hline
    \end{tabular}
    \caption{The cross sections after cuts of Eq.~\eqref{eq:basic-cuts} for backgrounds, in pb, and signals corresponding to flavon masses of 200, 1000, and 1400 GeV, in fb. The signal cross sections were calculated with $\Lambda=1$ TeV.}
    \label{tab:cut-efficiencies}
\end{table}
As we will show, the neural networks can efficiently distinguish between OS signals and backgrounds, while SS signals have very little contamination. Of course, the neural networks can also easily distinguish between OS and SS events. However, if we mix them together and frame the problem as OS+SS signals against backgrounds, we lose statistical significance as the number of SS signal events is much smaller than $t\bar{t}$ events even after pNN classification. Instead, we can calculate the statistical significance of the SS and OS signals separately after the pNN classification, combine them in quadrature, and profit from both channels. 

\subsection{Training and Validation of the parametrized Neural Networks}

 Our goal is to find the regions of the two-dimensional parameter space $m_a$ {\it versus} $c_{tq}/\Lambda$ where the HL-LHC can exclude, at the 95\% CL, or discover the flavons. We trained neural networks to separate signals and backgrounds in order to maximize the signal significance for a given mass hypothesis.
 The coupling just rescales the remaining number of events after the neural net separation, but the mass impacts the distribution of the kinematic variables of the signals. In principle, a neural net trained for a given flavon mass might perform poorly in identifying signals of other masses, what would force us to train a large number of algorithms to account for the change in the flavon mass parameter. Instead, it was demonstrated in~\cite{pnn1,pnn2} that a neural network parametrized by a small fixed grid of mass parameters can learn how to interpolate between the masses of the grid used to train the algorithm. In our case, we trained our pNN with signal events of the mass grid: 200, 500, 800, 1100, and 1400 GeV. We then used the model to classify events of other eight different masses. We checked that the pNN learns to distinguish the intermediate masses correctly by training dedicated neural nets for a couple of intermediate masses. The conditioning of the model to the flavon mass is done by adding a column to the dataset for the mass of the signal event. In the case of backgrounds, we randomly choose a mass value from a uniform distribution in the mass range of the grid.


\begin{figure}[ht]
    \centering    
    \includegraphics[width=0.45\linewidth]{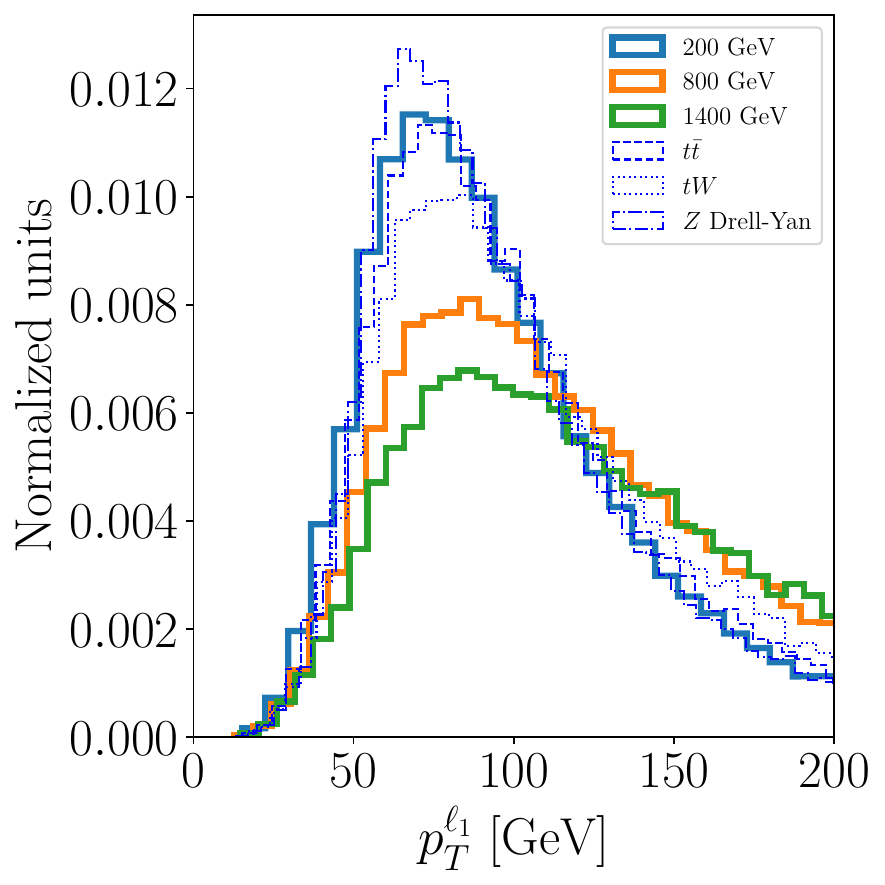}
    \includegraphics[width=0.45\linewidth]{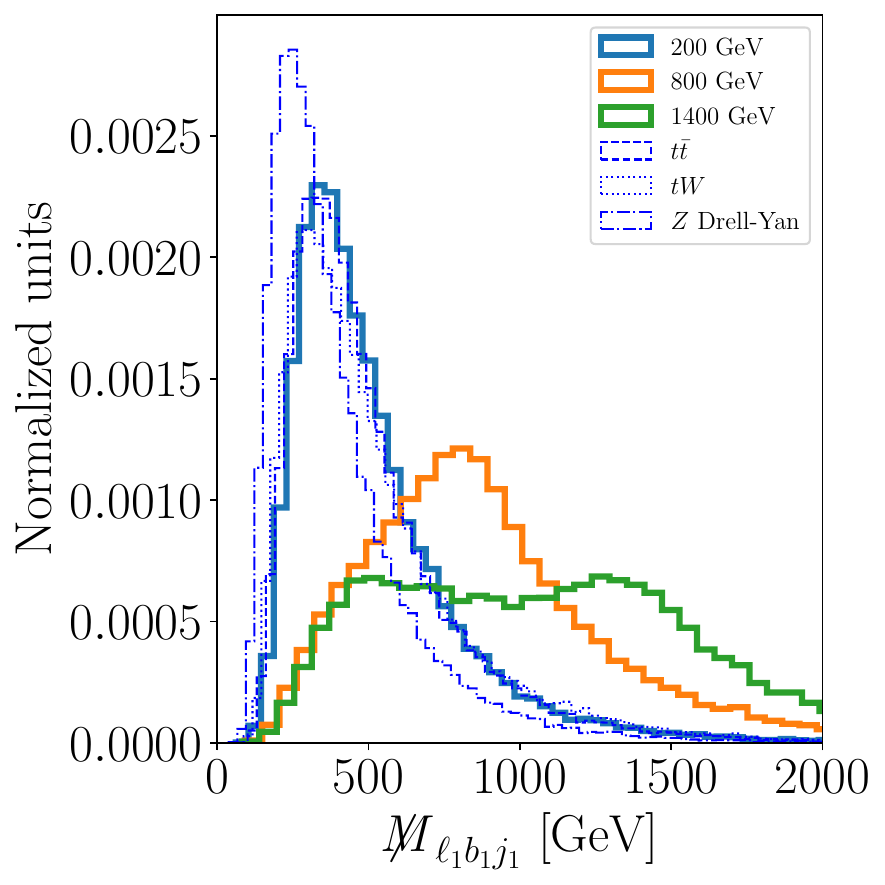}
    \includegraphics[width=0.44\linewidth]{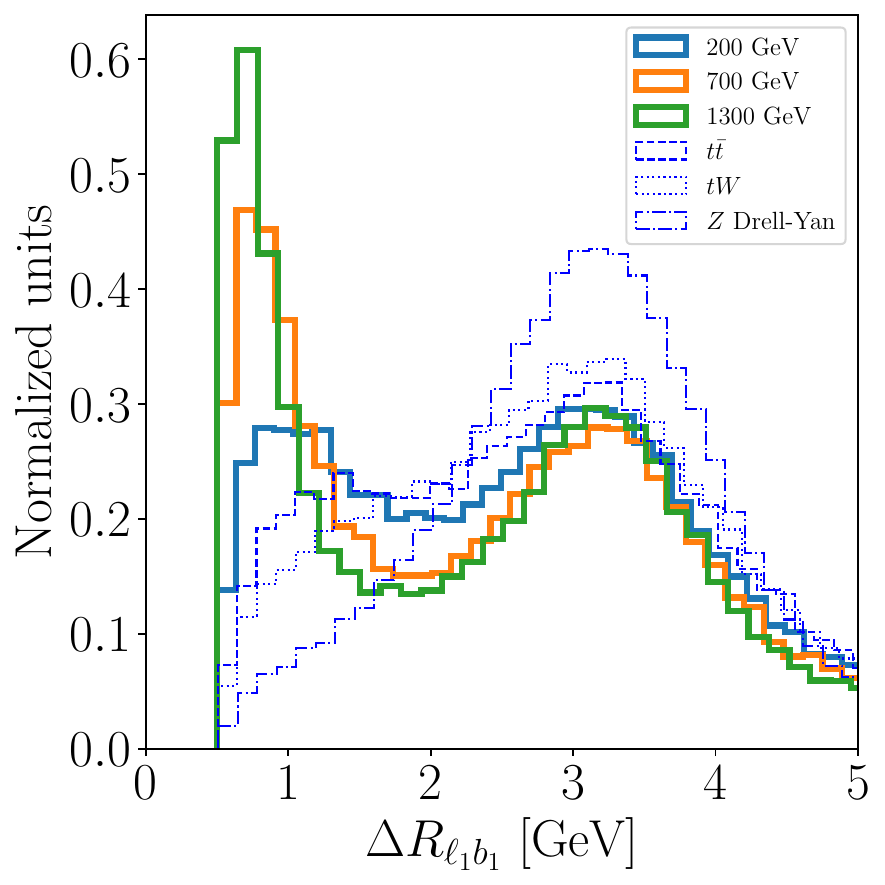}
    \includegraphics[width=0.45\linewidth]{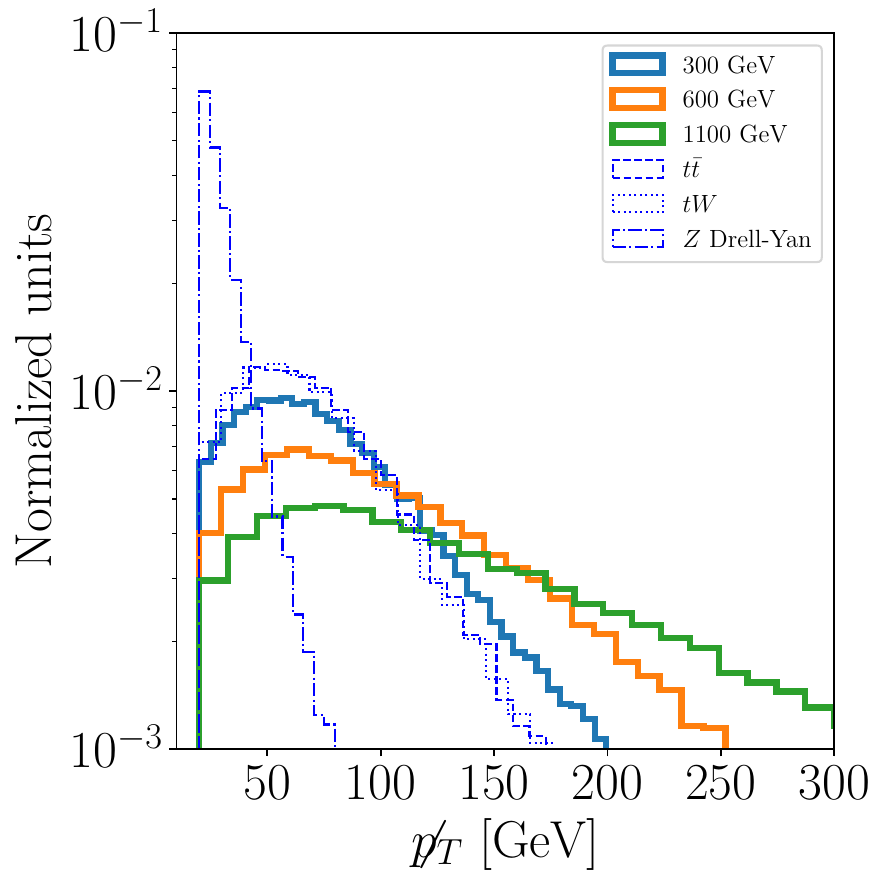}
    \caption{Examples of kinematic distributions used as high-level variables in the pNN training. For smaller $m_a$ all distributions are similar to the $t\bar{t}$ and $tW$ backgrounds. They become more distinct as $m_a$ grows, which partly explains why the pNN identifies signals with higher $m_a$ with greater accuracy.}
    \label{fig:distros}
\end{figure}

 The variables used to represent the events might facilitate the performance of the neural networks to separate the signal and background classes. Many high-level features can be constructed with the four-momenta of the leptons, bottom, and light-quark jets. We tested many configurations and found a very good representation of our events with the kinematic variables listed below:
\begin{itemize}
\item the transverse momentum of the two leading $b$-jets, the two leptons, and the two leading non $b$-jets: $p_{T_{\ell_{1, 2}}}, p_{T_{b_{1, 2}}}, p_{T_{j_{1, 2}}}$,
\item the missing transverse momentum, $\not\!\! p_T$,
\item the energy, at the laboratory center-of-mass frame, of the two leading $b$-jets, and the two leptons: $E_{\ell_{1,2}}$, $E_{b_{1, 2}}$,
\item the pseudorapidity of the two leading $b$-jets, the two leptons, and the leading non $b$-jet: $\eta_{\ell_{1,2}}$, $\eta_{b_{1,2}}$, $\eta_{j_1}$, 
\item the invariant masses: $M_{\ell_1 \ell_2}$, $M_{b_1 b_2}$, $M_{\ell_1 b_1}$, $M_{\ell_1 b_2}$, $M_{\ell_2 b_1}$, $M_{\ell_2 b_2}$, $M_{\ell_1 j_1}$, and ${\not\!\! M}_{\ell_1 b_1 j_1}$, the mass of the system $p_{\ell_1}+p_{b_1}+p_{j_1}+\not\!\! p$, where $\not\!\! p=(\sqrt{\not\!\! p_x^2+\not\!\! p_y^2},\not\!\! p_x,\not\!\! p_y,0)$ in the lab frame,  
\item the difference in the azimuthal angle of the vectors $p_{\ell_1}+p_{\ell_2}$ and $p_{b_1}+p_{b_2}$: $\Delta \phi_{\ell_1 \ell_2 b_1 b_2} = \Delta\phi_{\ell_1+\ell_2,b_1+b_2}$,
\item the invariant angular distances, $\Delta R_{i,j} = \sqrt{(\eta_i-\eta_j)^2+(\phi_i-\phi_j)^2}$: $\Delta R_{\ell_1 \ell_2}$, $\Delta R_{\ell_1 b_1}$, $\Delta R_{\ell_1 b_2}$, $\Delta R_{\ell_1 j_1}$, $\Delta R_{\ell_2 b_1}$, $\Delta R_{\ell_2 b_2}$, $\Delta R_{b_1 b_2}$,
\item the number of identified $\mu^+$, $\mu^-$, $e^+$, $e^-$, $b$-jets and non-$b$-jets in the final state.
\end{itemize}

Some of these variable distributions are shown in Fig.~\ref{fig:distros}.

Concerning the algorithm training, we separated the data into three parts: training, validation, and test sets. 
After every epoch -- the passing of the whole training data through the algorithm for updating its weights -- we checked the discerning power of the neural network by examining the training and validation metrics for overfitting.

A Bayesian-inspired optimization of neural network hyperparameters was done with the \texttt{Hyperopt} package \cite{hyperopt}. 
The hyperparameters considered were the initial learning rate, batch size, number of hidden layers, activation functions, dropout rate, and the number of neurons at each hidden layer. The best set we found is shown in Tab.~\ref{tab:hyperparam}. The minimization of the binary cross-entropy function was carried out with \texttt{Adam}~\cite{kingma2017adam}.
To avoid overfitting,  an early stopping criterion of 25 epochs without improvement in the loss function on the validation set was adopted with a total number of epochs set to 200. 

\begin{table}[t]
    \begin{tabular}{@{}cccccc@{}}
    \hline
    hidden layers & neurons & activation & drop-out & learning rate & batch size \\ \hline \hline
    3                      & 80               & ELU                          & 0.265             & 0.00156                & 80                  \\ \hline
    \end{tabular}
\centering
\caption{Hyperparameters and neural net architecture used to train the pNN. The hyperparameters were optimized with Hyperopt~\cite{hyperopt}.}
\label{tab:hyperparam}
\end{table}
To guarantee a reliable classification, a balance in the signal-to-background ratio in the samples of Monte Carlo events must be kept when training the pNN, according to Ref.~\cite{pnn2}. The training was performed with 270000 events evenly distributed among signals from the grid of masses and the three background sources. 

The performance of the pNN model was tested using a 10-fold cross-validation, splitting the training set containing backgrounds and the signals events corresponding to the grid masses into two datasets: one to train the pNN and another set held out to compute the model's accuracy. The mean accuracy across the 10 validation sets was 89\%.

\begin{figure}[t]
    \centering
    \includegraphics[width=0.45\linewidth]{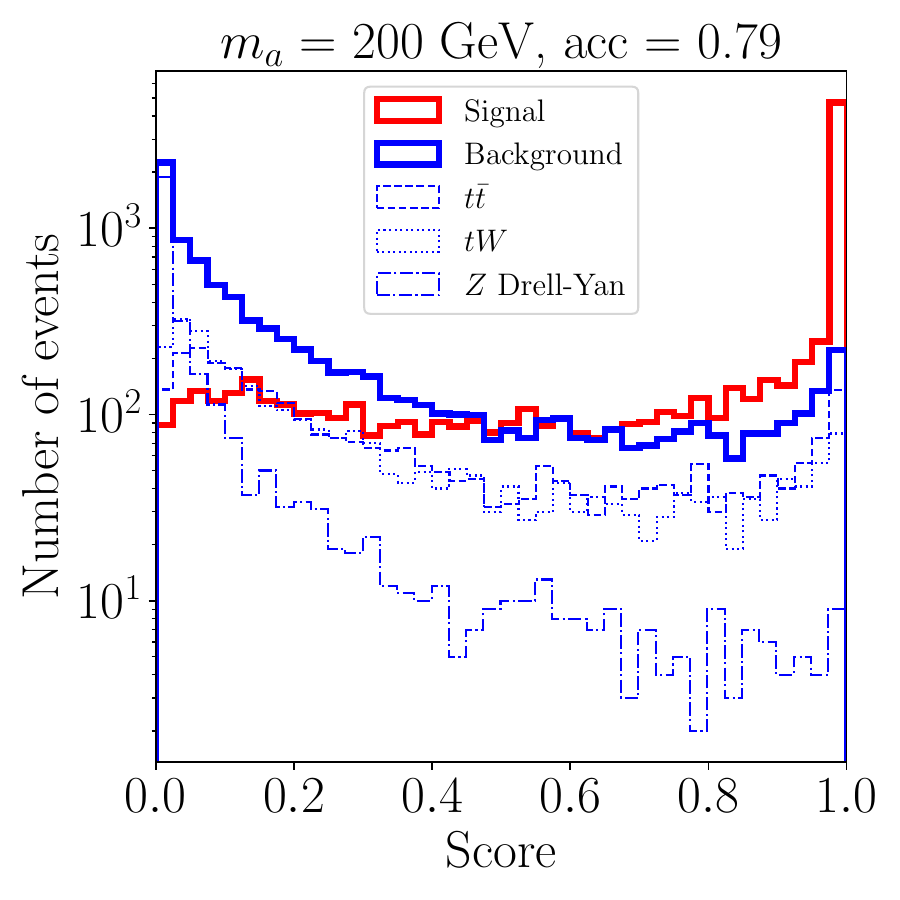}
    \includegraphics[width=0.45\linewidth]{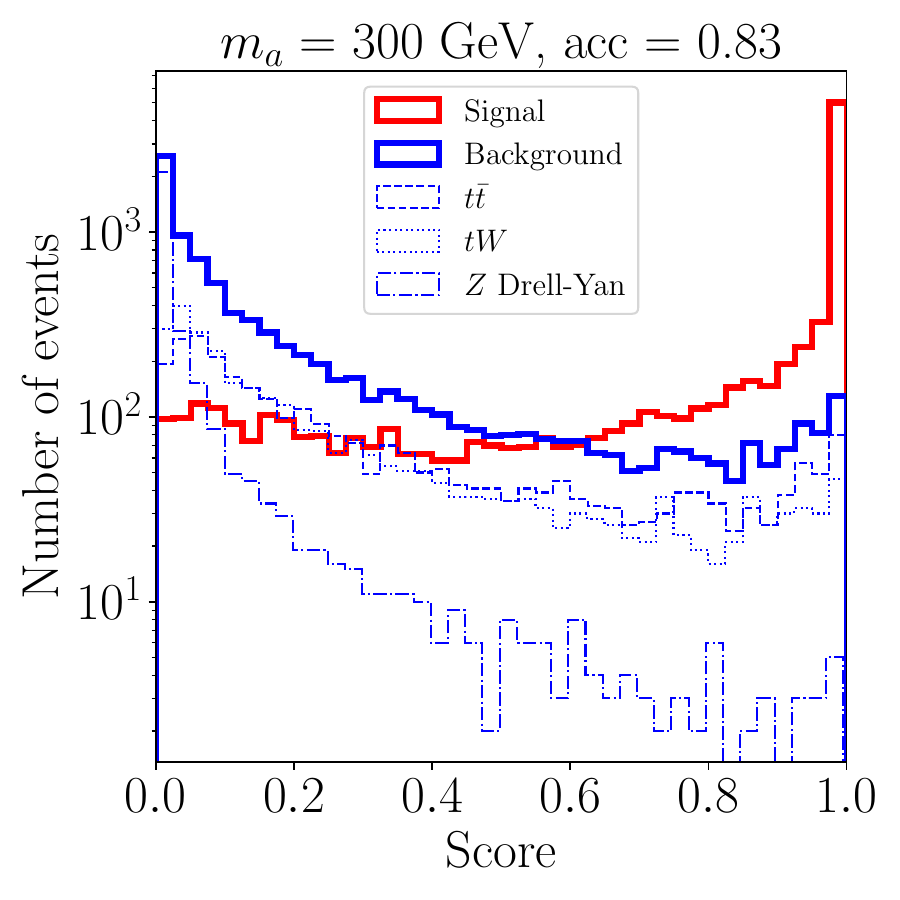}
    \includegraphics[width=0.45\linewidth]{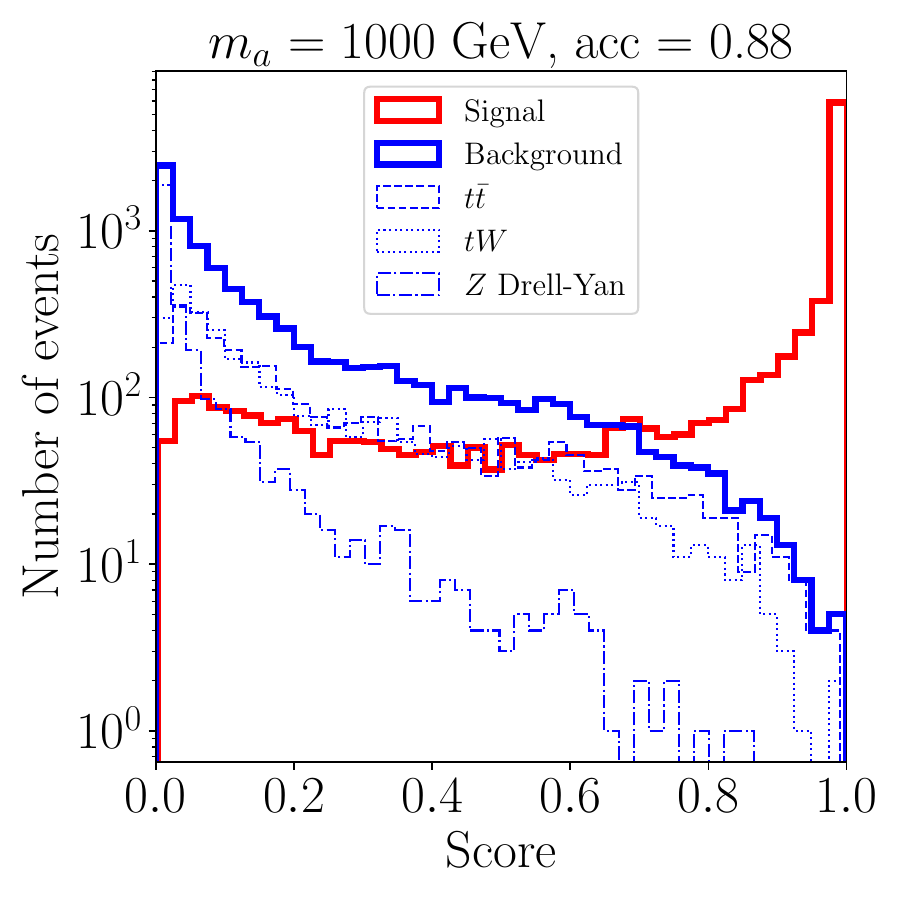}
    \includegraphics[width=0.45\linewidth]{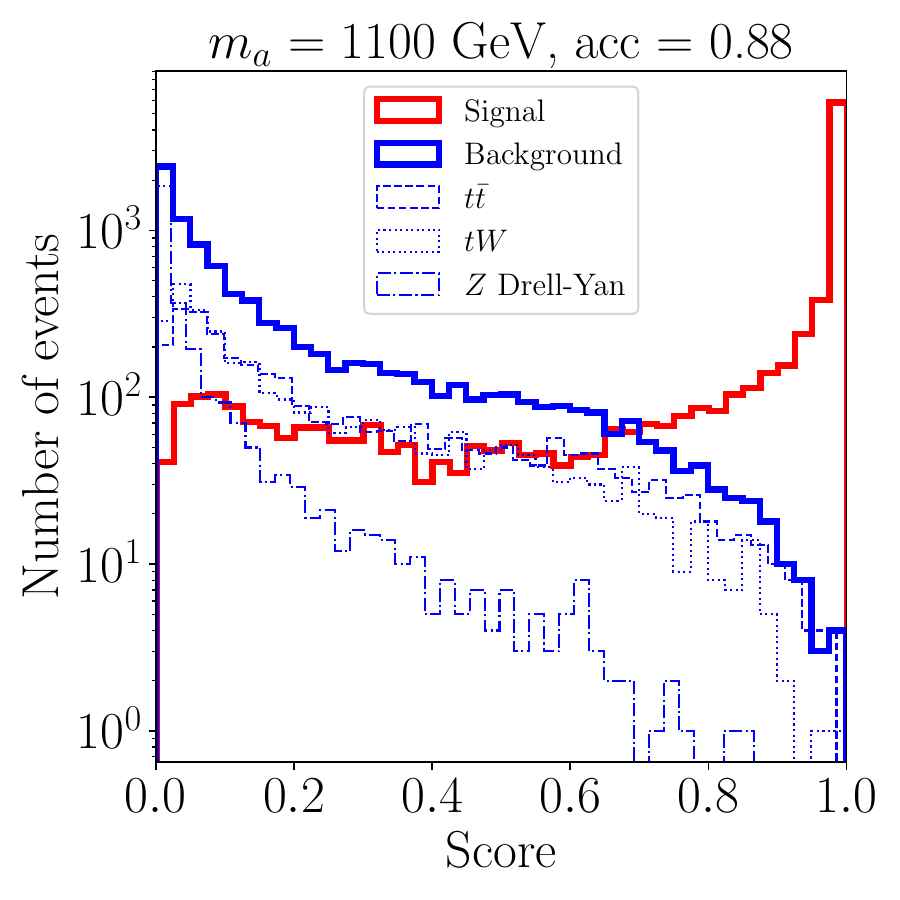}
    \caption{Output scores of signal(red) and background(blue) on the pNN classification task. The higher flavon masses are more easily distinguished once their kinematic distributions get more distinct from the backgrounds. We also show the performance of the algorithm for each background separately. We also show the corresponding accuracy (acc), defined as the fraction of correct identifications in the sample.}
    \label{fig:scores}
\end{figure}

After training and cross-validation, we used the best cross-validation model in the test data, which contains signal events with masses from 200 to 1400 GeV in steps of 100 GeV, 9000 events each, and 9000 background events in total. The test set contains signal events with masses absent in the training set. Nonetheless, the performance of the pNN on these events is almost as good as on the grid masses used for training.

\begin{figure}[t]
    \centering    
    \includegraphics[width=0.45\linewidth]{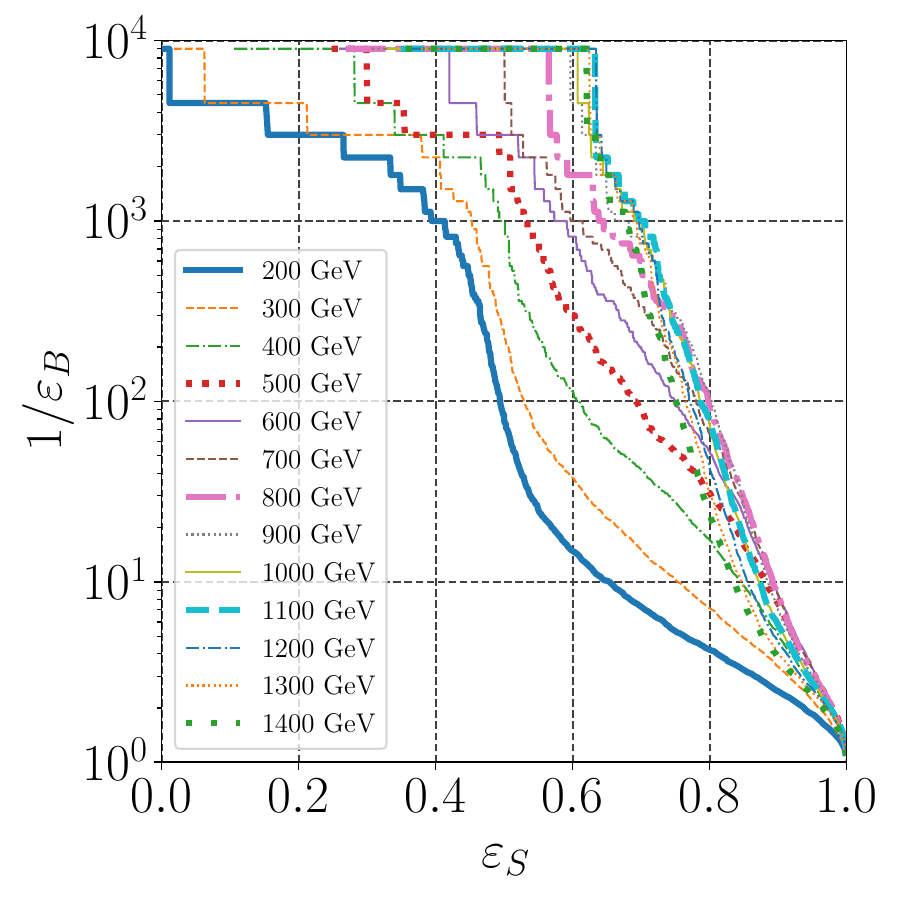}
    \caption{The background rejection, $1/\varepsilon_B$, against the signal efficiency, $\varepsilon_S$, after cutting on the pNN output scores.}
    \label{fig:roc}
\end{figure}

The neural network predicts an output score corresponding to the chance of an event being tagged as a signal. Fig.~\ref{fig:scores} shows the distribution of the signal (red) and background (blue) events as functions of the classifier output score. Lighter flavons are more difficult to discriminate from the background, as could have been anticipated from the distributions in Fig.~\ref{fig:distros}. In Fig.~\ref{fig:roc}, we depict the algorithm performance for each test mass in terms of the signal efficiency, $\varepsilon_S$, and background rejection, $1/\varepsilon_B$, for events that survive a given threshold score cut. For a fixed signal efficiency, the background rejection increases from 200 GeV flavons to 1000 GeV, reaching a maximum and saturating, showing no improvement for larger masses. The neural network presents an exquisite discerning power, keeping only $\sim (10^{-3})10^{-4}$ of background events for a signal efficiency of 50\% in the (worst)best mass cases. The accuracy (acc) also increases from light to heavy flavon masses as can be seen from Fig.~\eqref{fig:scores}.

%

\section{RESULTS}
\label{sec:results}
\begin{figure}[t]
    \centering    
    \includegraphics[width=0.45\linewidth]{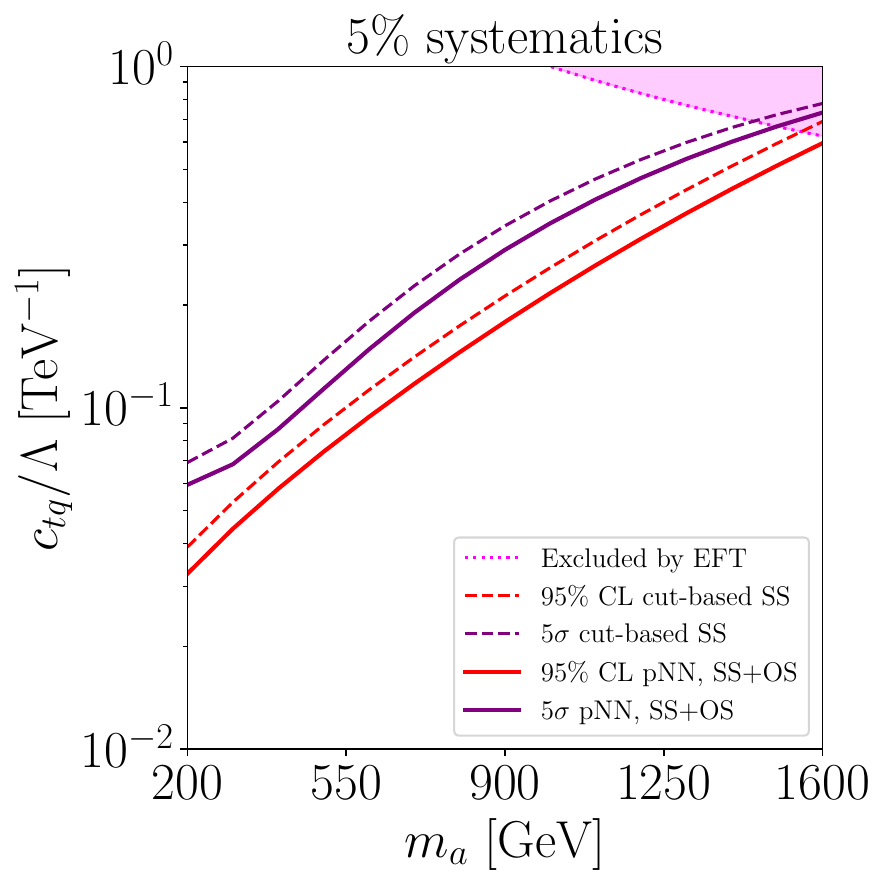}
     \includegraphics[width=0.45\linewidth]{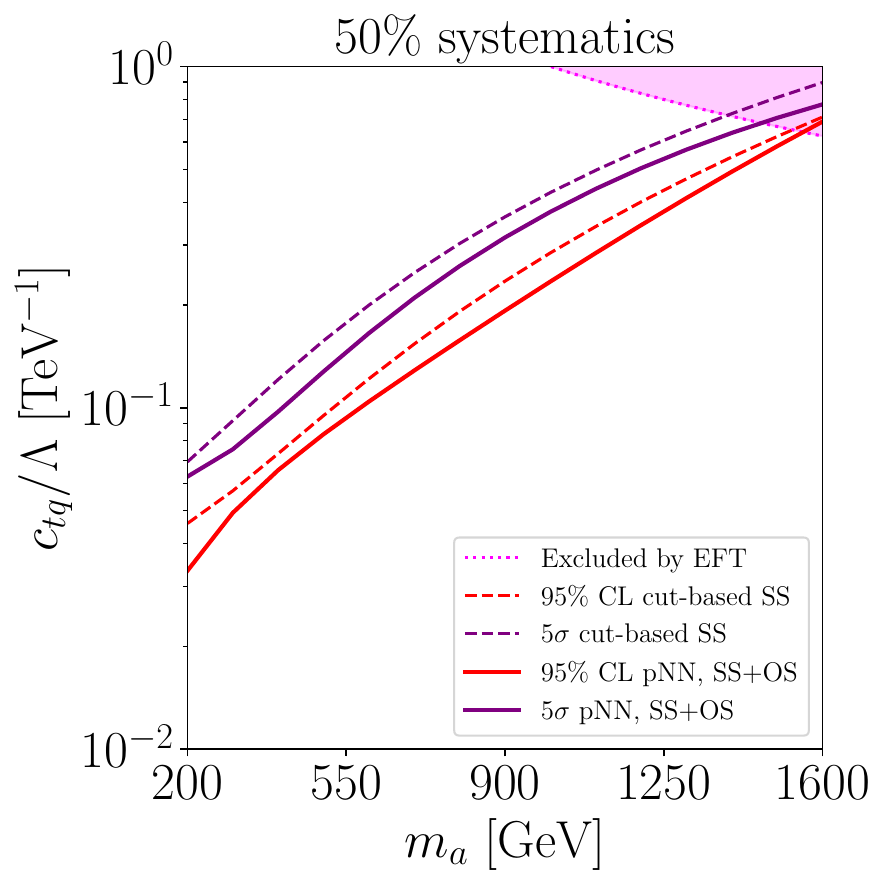}
    \caption{The red(purple) lines represent the 95\%($5\sigma$) CL limits on the coupling and flavon mass parameters for a luminosity of 3 ab$^{-1}$. The dashed(solid) lines depict limits based on a cut-and-count on SS signals (pNN on SS+OS signals) analysis. The shaded upper right region delimits the region where EFT is valid. At the left(right) panel, the systematic uncertainty in the background normalization is 5\%(50\%).}
    \label{fig:result}
\end{figure}

 Because the background contamination in SS samples is expected to be rather low, we first performed a simple cut analysis requiring two same-sign leptons in the event alongside the cuts of Eq.~\eqref{eq:basic-cuts}. In this case, we avoid the large backgrounds to opposite-sign leptons at the cost of halving the number of signals. The strategy pays off from the perspective of a cut-based analysis, and the results for 95\% CL and $5\sigma$ discovery regions in the coupling {\it versus} mass plane are shown in Fig.~\ref{fig:result}. The statistical significances were calculated using the Asimov formula, assuming systematic uncertainties in the background rates~\cite{Cowan_2011} assuming an integrated luminosity of 3 ab$^{-1}$. Notice the stability of the SS cut-based results against systematic uncertainties due to the large $S/B$ ratio after cuts.

 The neural network allows us to efficiently distinguish between signals and background irrespective of their lepton charges. The charges of leptons can be used later to separate the events into OS and SS classes to calculate their statistical significances separately and then combine them in quadrature, raising the overall significance compared to the SS significance alone.

 After training the pNN, as described in the previous section, we determined the number of surviving signals and background events by placing cuts on the test set scores. The cut on the pNN score was scanned to achieve the highest SS and OS statistical significances, provided they satisfy the minimum number of events compatible (2 events for 95\% CL exclusion and 25 events for 5$\sigma$ discovery). The final significance of the signal is the quadrature of both SS and OS significances, as discussed.

 We plot the resulting 95\% CL exclusion and $5\sigma$ discovery regions for pNN classification in Fig.~\ref{fig:result} also assuming an integrated luminosity of 3 ab$^{-1}$. The pNN approach probes higher cutoff scales than the SS cut-based approach for each mass utilizing both SS and OS events. Roughly speaking, pNNs can probe couplings 50\% smaller than the cut-based SS strategy. The advantage faints as the flavon mass increases, but it is still better up to 1.6 TeV mass, where the EFT validity is expected to be broken. The region $m_a > \Lambda$, highlighted in magenta in Fig.~\ref{fig:result},  refers to the validity limit of the EFT. Effective couplings of order $10^{-2}$ TeV$^{-1}$ can be probed for flavon masses up to $\sim 600$ GeV.

 Another important feature of our results is the reduced sensitivity to systematic uncertainties on the background rates. As we see in Fig.~\ref{fig:result}, the limits barely change from the 5\% to 50\% systematics scenario due to the exquisite job of the neural network in separating signals from the background with high signal-to-background ratios. These results extend those for $m_a\lesssim 200$ GeV where the flavon prefers to decay into tau leptons~\cite{Alves:2023sdf}. In combination, they cover the heavy flavon mass range from 10 to 1600 GeV, complementing the results for lighter flavon masses, $m_a\lesssim 10$ GeV~\cite{Cornella:2019uxs}.
 

\section{CONCLUSIONS}
\label{sec:conclusions}
 Flavor-violating phenomena at colliders are an interesting area for new physics searches. Contrary to light quarks and leptons, flavor-violating interactions involving top quarks and up or charm quarks are much less constrained by searches for flavor-changing neutral currents. Models like the Froggatt-Nielsen mechanism, which aims to explain the Yukawa hierarchy, predict the existence of an axion-like particle that can mediate such flavor-violating interactions, which we call flavons. We assume these scalars predominantly change the flavor of the fermions they interact with, avoiding constraints from flavor-conserving searches.

 This work explored a distinctive phenomenology of flavons at the 14 TeV LHC. Pair and single flavon production, plus virtual $t$-channel contributions, give rise to pairs of opposite and same-charge top quarks. The decay of same-charge tops to same-sign leptons leads to almost background-free signals at the expense of halving the number of signal events. A simple cut-and-count strategy is able to probe flavon couplings of the order of $10^{-2}$ TeV$^{-1}$. Instead of wasting all the opposite-charge top quarks, we trained a neural network to clean up the signals from $t\bar{t}$ and other backgrounds. We chose to work with parameterized neural networks to get a smooth classifier as a function of the flavon mass. After the classification, we separated the samples according to the lepton pairs charges and combined the SS and OS signal significances in quadrature, thus profiting from all flavons produced. The combined analyses extended the limits by 50\% compared to the cut analysis based on SS tops only. Moreover, the combined analysis turned out to be very insensitive to systematic uncertainties in background rates. 

 With 3 ab$^{-1}$, we estimate that the 14 TeV LHC can exclude 200(1600) GeV flavons for couplings as weak as $c_{tq}/\Lambda \sim 3(50)\times 10^{-2}$ TeV$^{-1}$. The discovery prospects are also good, 200(1600) GeV flavons can be discovered for couplings $c_{tq}/\Lambda\gtrsim 6(70)\times 10^{-2}$ TeV$^{-1}$. Our findings extend previous searches for light flavons up to 10 GeV~\cite{Cornella:2019uxs}, and heavy flavons with masses up to $\sim 200$ GeV where they might prefer to decay into tau leptons~\cite{Alves:2023sdf}.

\bigskip{}
	
\textbf{Acknowledgments}: This study was financed in part by Conselho Nacional de Desenvolvimento Científico e Tecnológico (CNPq), via the Grants No. 307317/2021-8 (A. A.) and No. 308280/2023-7 (A. G. D.), and in part by the Coordenação de Aperfeiçoamento de Pessoal de Nível Superior – Brasil (CAPES) – Finance Code 001 (D. S. V. G.). A. A. also acknowledges support from the FAPESP (No. 2021/01089-1) Grant. E. d. S. A. partially thanks FAPESP for its support in the initial steps of this work (Grant  No.  2022/07185-5). 

\bibliography{myrefs}

\end{document}

%% file: qg_at_T.tex
\begin{tikzpicture}[baseline=(current bounding box.center)]
    \begin{feynman}
        \vertex (a2);
        \vertex [right=of a2] (a3) {\(a\)};
        \vertex [left=of a2] (a1) {\(q\)};
        \vertex [below=of a2] (b2);
        \vertex [right=of b2] (b3) {\(t\)};
        \vertex [left=of b2] (b1) {\(g\)};
        \diagram* {
        (a1) -- [fermion] (a2) -- [fermion, edge label=\(t\)] (b2) -- [gluon] (b1),
        (a2) -- [scalar] (a3),
        (b2) -- [fermion] (b3)
        };
    \end{feynman}
\end{tikzpicture}

%% file: qq_aa.tex
\begin{tikzpicture}[baseline=(current bounding box.center)]
    \begin{feynman}
        \vertex (a2);
        \vertex [right=of a2] (a3) {\(a\)};
        \vertex [left=of a2] (a1) {\(q\)};
        \vertex [below=of a2] (b2);
        \vertex [right=of b2] (b3) {\(a\)};
        \vertex [left=of b2] (b1) {\(\bar{q}^{\prime}\)};
        \diagram* {
        (a1) -- [fermion] (a2) -- [fermion, edge label=\(t\)] (b2) -- [fermion] (b1),
        (a2) -- [scalar] (a3),
        (b2) -- [scalar] (b3)
        };
    \end{feynman}
\end{tikzpicture}

%% file: at.tex
\begin{tikzpicture}[baseline=(current bounding box.center)]
    \begin{feynman}
        \vertex (a2);
        \vertex [right=of a2] (a3) {\(t\)};
        \vertex [left=of a2] (a1) {\(q\)};
        \vertex [below=of a2] (b2);
        \vertex [right=of b2] (b3) {\(t\)};
        \vertex [left=of b2] (b1) {\(\bar{q}^{\prime}\)};
        \diagram* {
        (a1) -- [fermion] (a2) -- [scalar, edge label=\(a\)] (b2) -- [anti fermion] (b1),
        (a2) -- [fermion] (a3),
        (b2) -- [fermion] (b3)
        };
    \end{feynman}
\end{tikzpicture}